# A new analysis method to determine β-decay half-lives in experiments with complex background[*]

T. Kurtukian-Nieto[a,†], J. Benlliure[a], K.-H. Schmidt[b]

[a]*Universidad de Santiago de Compostela, E-15782 Santiago de Compostela, Spain,*
[b]*Gesellschaft für Schwerionenforschung mbH, D-64291 Darmstadt, Germany*

**Abstract**
This paper reports the first application of a new technique to measure the β-decay half - lives of exotic nuclei in complex background conditions. Since standard tools were not adapted to extract the relevant information, a new analysis method was developed. The time distribution of background events is established by recording time correlations in backward time. The β half lives of the nuclides and the detection efficiency of the set-up are determined simultaneously from a least-squares fit of the ratio of the time-correlation spectra recorded in forward and in backward time, using numerical functions. The necessary numerical functions are calculated in a Monte-Carlo code using the known operation parameters of the experiment and different values for the two free parameters, half-life and detection efficiency, as input parameters.

## 1. Introduction

The half-life is a fundamental property of radioactive nuclei, carrying important information on their intrinsic structure. For the β decay, the situation is particularly complex, because it may populate a large number of levels in the daughter nucleus. Theoretical predictions are still rather uncertain, and therefore experimental work on this field is most important. This is even more the case, since β decay plays a decisive role in several astrophysical processes. A prominent example is the r-process nucleosynthesis [1,2], which relies on the β-decay properties of very neutron-rich nuclei along the r-process path. The present work deals with a measurement on β half-lives of nuclei close to the 126-neutron shell south of lead. This region is of exceptional interest to understand how the nuclear structure evolves on the neutron-rich side beyond the doubly-magic nucleus $^{208}$Pb, which is rather well understood in terms of the shell model. There has been little progress for nuclei in this region in the past due to the difficulty in producing and identifying these nuclides. This is in contrast to lighter elements, which are well populated by fission of actinides. The main reason for the progress achieved in the present experiment was the choice of a novel production mechanism: The nuclides of interest were produced by cold fragmentation using relativistic heavy-ion beams [3].

---

[*] This work forms part of the PhD Thesis of T. Kurtukian-Nieto
[†] Corresponding author, e-mail address: kurtukia@cenbg.in2p3.fr

The nuclei of interest were available as secondary projectiles. They were identified in-flight and implanted into an active catcher. The basic information for determining their half-lives, the implantation times and their subsequent decays were registered. Due to the time structure of the beam from the SIS18 heavy-ion synchrotron, the implantations, the beta decays and background events from several sources are influenced by the repetition rate and the length of the beam pulses.

A number of well established methods for the extraction of half-lives from measured raw data exist, from which the one which is best suited for the specific nature of the data as determined by the experimental approach may be chosen. Unfortunately, we found that none of these established methods are suited for the specific conditions of the present experiment. The complex time structure of the registered events due to the periodic operation conditions of the synchrotron accelerator, which provided the relativistic heavy-ion beam of our experiment, required the development of a new mathematical analysis method.

In this paper we will give a short review on the characteristics of the raw data provided by different experimental approaches and on the essential conditions and parameters of the well established analysis methods. This survey will demonstrate the need for the development of a new mathematical model for extracting the half-lives of the nuclear species produced in the present experiment. The following chapters will give a detailed documentation of the new analysis method we have developed and investigate its strengths and limitations. The individual steps of the analysis are presented and its success is demonstrated.

## 2. Experimental approaches for measuring half lives

The technique of choice for measuring half-lives will depend on the lifetimes themselves and on the production of the nuclei investigated. There are essentially two typical classes of experimental conditions for measuring half-lives[‡]:

### 2.1 Activation method
In the first type of experiment, it is possible to produce a large number of nuclei in a given time period, which is short compared to their half-life. This is the case if the nuclide in question can be produced with high cross-section. In this case, the experiment consists of an activation phase and a recording phase. These phases might be repeated several times. After the activation time, the decrease of the number of nuclei in the sample is registered. This can be done either by determining the decay rate or by counting the number of nuclei still present. These two quantities are coupled by the continuity condition.

---

[‡] We only consider experimental approaches which rely on directly recording the times of individual events. Exploiting indirect signatures of decay times like Doppler shift of gamma rays during stopping the nuclei of interest are beyond the scope of the present paper.

There exist some variants of the activation-type experiment. First, the decays are registered; secondly, the number of nuclei still present in the probe is determined as a function of time. In both cases, the spectrum follows an exponential function. For very long half-lives, the decay rate is approximately constant during the experiment, and the value of the half life can be obtained directly from the ratio of the number of detected decays per time unit over the total number of nuclei in the sample. Background may consist of nuclei with other half-lives (contaminations, daughter nuclei) or a constant rate of parasitic events.

**2.2. Delayed-coincidence method**
Another type of experiment is characterised by short half lives and/or low production rates. If the rate of produced nuclei is lower than the inverse of its half life, it is convenient to record the time of production and look for the time of the consecutive decay. These time differences are sorted into a spectrum.

In this case, there are roughly three different variants on recording the data: One may record all events, which look like a decay, after any produced nucleus, or one may record all events, which look like a decay, after the last produced nucleus, or one may record only the first event, which looks like a decay, after the last produced nucleus.

In 'simple' experimental conditions, the average production rate is constant. That means that there is no time structure in the production rate. Independently from the recording options mentioned above, the 'true' correlations follow an exponential distribution. Also in delayed-coincident experiments, background may exist, consisting of nuclei with other half-lives, decays of daughter nuclei, and a constant rate of parasitic events. A constant background rate appears as a constant rate if all events are registered, but as an exponential distribution if the time scale refers only to the last produced nucleus.

The condition that the rate of produced nuclei is lower than the inverse of its half life should consider an eventual position sensitivity of the experiment. If the detector, which registers the produced nuclei and the decays, is subdivided, the situation is equivalent to a number of independent experiments with lower rate, corresponding to the granularity of the detector.

The background conditions of the experiment are very much influenced by how well the nuclei as well as their decay are specific for a certain decay process. Best conditions are met if the nuclei are completely identified in mass and atomic number and if the decay is unambiguously identified, e.g. by its energy lines in alpha and gamma decay, using a high-resolution detector. In this case, the decays of many different species can be investigated simultaneously without any interference. Such an experiment is equivalent to a number of experiments with different nuclear species. Still the identification of either the nucleus or the decay is helpful for improving the experimental conditions, mostly in terms of background conditions. Since the β spectrum has the property of being continuous, β-decay is not specific enough to be attributed to a specific nucleus, increasing the probability of a random correlation.

**2.3. Delayed coincidence method with additional time structure**
The experiment considered in the present work falls into the category of delayed coincidences. However, as already mentioned, the source intensity is not constant. It is modulated with a periodic function due to the extraction cycles of the synchrotron. In this case, the recorded time-correlation spectra do not directly reflect the decay properties of the nuclei involved but also show signatures of the time structure of the beam.

# 3. Standard analysis methods to extract nuclear half-lives

The standard analysis methods used to extract nuclear life times rely on the analytical formulation of the time distributions and on appropriate tools to extract the relevant parameters of the corresponding analytical function from the measured data. A short overview on these methods is presented in order to explain and justify the needs for a new approach, adapted to the data measured in the specific conditions of the present experiment.

The radioactive decay is a random process, characterised by a certain decay probability per time $dP/dt = \lambda$. As all radioactive nuclei decay independently from the others, the density distribution of radioactive decays of one species of nuclei, is expressed by the exponential law

$$\frac{dn}{dt} = n_0 \cdot \lambda \cdot e^{-\lambda t}$$

where $n_0$ is the initial number of radioactive nuclei. Thus, $n(t) = n_0 \cdot e^{-\lambda t}$ is the number of nuclei remaining at time $t$.

The statistical analysis to deduce the decay constant $\lambda$ can be a complicated task since the radioactive decays can only be observed in a limited time range, and in addition, events of other species which decay with different decay constants or background events may be mixed in. Also if daughter nuclei produced in the primary decay are also radioactive, an even more complex situation appears.

Several methods have been developed to extract half-lives of radioactive species from experimental data and to determine their statistical uncertainties. Some of the most frequently used approaches will be mentioned below.

**3.1 Mean value**
If the sample under study contains only one species of radioactive nuclei and any background is excluded, the first moment or average of the measured decay times is the best approximation to the inverse of the decay constant, $1/\lambda$ and the mean relative error is $1/N$, being $N$ the total number of decays. [4]

$$\bar{t} = \sum_{i=1}^{n} t_i = \frac{1}{\lambda} = \tau = {T_{1/2}}/{\ln 2}$$

In order to be able to apply this method, the full time range must be covered by the measurement, that is, the minimum time must be very small, and the maximum time must be very large compared to $1/\lambda$.

**3.2 Decay curve**
In the approach, mostly applied to more complex radioactive-decay data, the individual decay times are sorted into a spectrum with time intervals of constant length $\Delta t$. The distribution for the occurrence of independent events in a given time interval follows the Poisson statistics but for larger number of counts it can well be approximated by a Gaussian distribution.

As mentioned above, the expected shape of the time-interval spectrum starting from any time is given by an exponential function, and the decay constant can be determined by a fit. A least-squares fit is restricted to large event numbers, while the maximum-likelihood method can also be applied in the case of poor statistics [5]. Contributions of other radioactive species and of background events can be recognised and extracted by using a more complex fit function, being the sum of several exponential distributions or more complex functions and eventually a constant background value.

It has also been proposed to sort the individual times into a spectrum with time intervals $\Delta t$, which have a width that is proportional to the time $t$, that means $\Delta t/t$ = constant [6]. This logarithmic time-bin presentation facilitates the visualisation of the decay-time spectra in case of very few events and allows the storing of the relevant information of decay times over a large range with a moderate number of channels. Also in this presentation, the decay constant can be determined from a fit with an analytical function, which is, however, bell-shaped and slightly asymmetric with a maximum located at $\tau$ [4,6].

# 4. Experiment

The experiment, which motivated the present piece of work, was focused on the measurement of the β-decay half-lives of heavy neutron-rich nuclei. The SIS18 accelerator of the Gesellschaft für Schwerionenforschung (GSI) provided the relativistic primary beam, and the high-resolution magnetic spectrometer FRS [7] allowed the produced nuclei to be identified in mass and atomic number.

A beam of $^{208}$Pb at 1 $A$ GeV impinged upon a beryllium target. The beam intensity was $10^7$ ions/s, the spill length was 2 s and the repetition cycle 10 s. As a result of the collision, the target nucleus will abrade nucleons from the initial projectile, producing both stable and radioactive secondary beams in flight. Heavy neutron-rich isotopes from $Z=70$ up to $Z=83$ lying at and beyond the limit of the previously known nuclei have been identified. Production rates of exotic nuclei are in general very low and, as a consequence, the experimental technique chosen to measure half-lives was the delayed-coincidence technique. In order to perform the half-life measurements, it is necessary first to identify each fragment without any ambiguity. Then, the selected ions are stopped and finally their β-decay is measured, correlating the delayed βs with the parent nuclei.

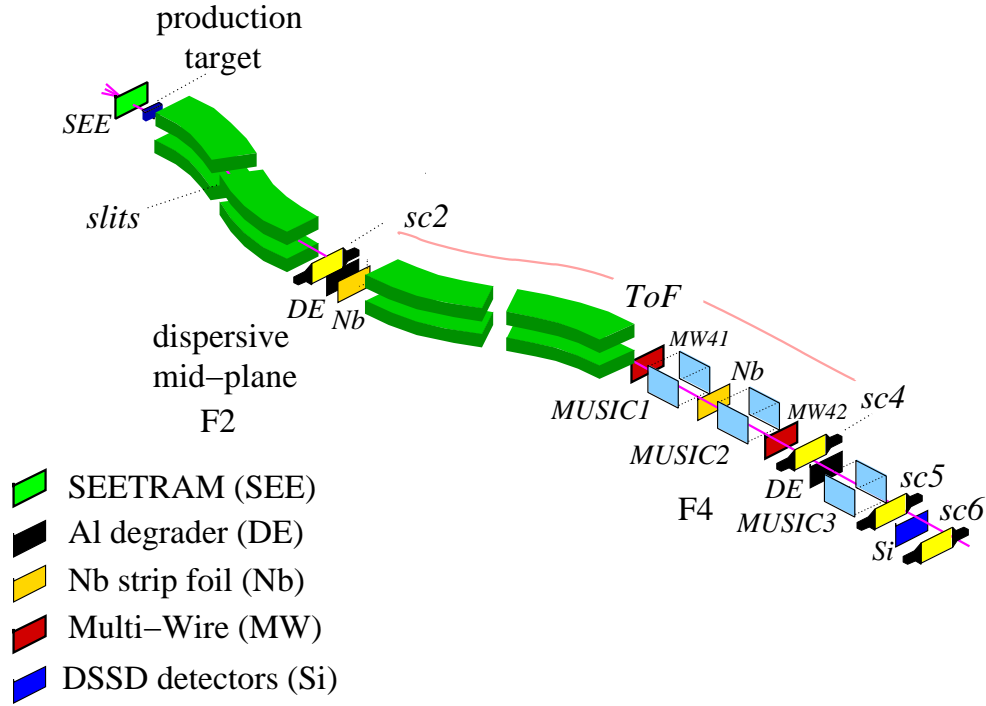

Figure 1: Experimental setup

The magnetic selection of the heavy neutron-rich nuclei was performed with the fragment separator FRS, schematically shown in Figure 1. The first step in the separation was achieved by the first two dipoles. Since the reaction mechanism approximately conserves the velocity, this selection is mainly sensitive to the ratio of mass number and ionic-charge state $A/q$. The magnetic fields of the first two dipoles can be opportunely tuned so that a selected nucleus $(A,Z)$ transverses the FRS following the central trajectory. Nuclei with close values to $A/Z$ will also be transmitted.

An additional selection is achieved by introducing a velocity degrader in the intermediate dispersive focal plane. The fragments transmitted by the first two dipoles with close values of $A/Z$, will have different momentum values after the reduction of their velocity, according to the energy loss in the degrader. Then, they can be separated in the second stage of the FRS by a new selection in magnetic rigidity. As a consequence, by tuning the ion optics, it is possible to select just a small region of the chart of nuclides. The range differences of the transmitted nuclei can also be exploited to achieve a further selection at the implantation stage. This allows experiments to be carried out focusing either on one nucleus or on a 'cocktail' with a controlled background.

The nuclei were identified in mass and atomic number by means of the measurements of the magnetic rigidities, time-of-flight (ToF) and energy loss of each fragment passing through the FRS. By slowing down the ions with an aluminium homogeneous degrader, it is possible to implant the ions in an active stopper, a highly-pixelated Si detector stack which allows for the correlation in time and space of the signal from the implanted ion with the subsequent signal produced by the β-decay.

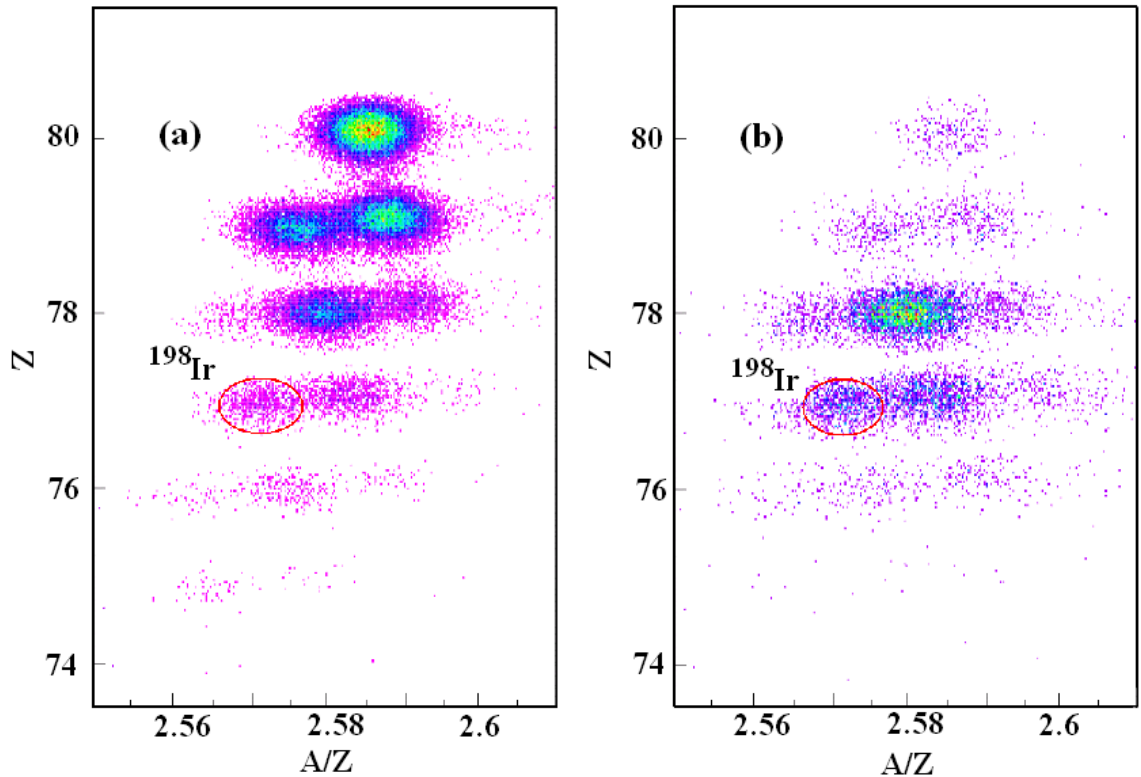

Figure 2: Particle identification plots corresponding to a FRS setting optimised to transmit $^{198}$Ir with a mono-energetic degrader. The figure shows the total production yield at the end of the FRS (a) and the ions which were implanted in the 1mm-thick DSSD (b).

Between 6 and 9 different nuclides transmitted to the final focal plane were implanted into the active stopper (see Figure 2), which allows us to measure the half-life of several species at the same time. Fragments with shorter ranges were stopped either in the degrader or in the layers of matter before the active stopper. The veto scintillator (SC6) rejects more penetrating or secondary fragments produced in any of the layers of matter.

The active stopper used in the experiment consists of a stack of 4 Micron Semiconductor Ltd [8] W(DS)-1000 5 cm x 5 cm double-sided silicon strip detectors of 1 mm thickness, providing enough depth of silicon for the implantation of several species and for the observation of the β decays. The readouts of these detectors are divided into 16 strips of 3.12 mm pitch in both the front and back side. The high segmentation of the detectors leads to a situation which is equivalent to a number of independent experiments with lower rate.

The use of a monoenergetic degrader [9] at the FRS provides a horizontal dispersion and a narrow range distribution of fragments in the active stopper. The narrow range of fragments increases the implantation efficiency and allows several exotic nuclei to be caught in a thin active stopper (< 1 mm). The horizontal dispersion allows us to take advantage of the high pixelation of the active stopper to reduce the probability for multiple implantation of nuclei in the same pixel.

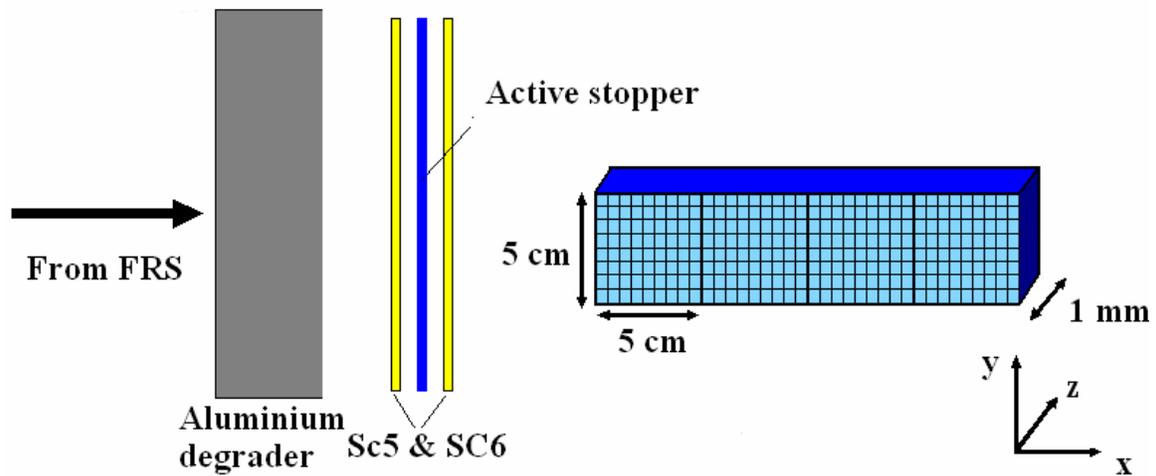

Figure 3: Detection setup for fragment – β-time correlations.

In order to implant a given nuclide in the middle of the active stopper, we calibrated first the thickness of the degrader by using the primary beam, $^{208}$Pb. In Figure 4 we show the energy loss of the beam in the scintillators in front (SC5) of and behind (SC6) the active stopper for different degrader thicknesses. In cases 'A' and 'B', the energy loss in the scintillator in front of the active stopper is proportional to the energy loss in the scintillator behind it. This corresponds to degrader thicknesses thin enough to allow the beam to cross both scintillators. As we progressively increase the degrader thickness, the scintillator behind the active stopper, SC6, starts to prevent the beam from passing through. The energy loss in that scintillator starts to decrease, since the projectile deposits less energy due to the decreasing range in the scintillator. This effect can be seen in regions C, D, E, and F in Figure 4. Region G corresponds to a degrader thickness for which half of the beam distribution reaches the scintillator SC6. If we add to this degrader, the aluminium-equivalent thickness of the layers of matter that separates the scintillator SC6 and the active stopper, and half of the thickness of the DSSD, we would then implant the beam right in the middle of the active stopper. This degrader thickness was used to calibrate the implantation of different species, so that

$$T(X) = R(X) - R(^{208}Pb) + T(^{208}Pb)$$

where $T(X)$ is the required degrader thickness to implant a given nuclide which has a range $R(X)$ at the exit of the FRS, in the middle of the active stopper.

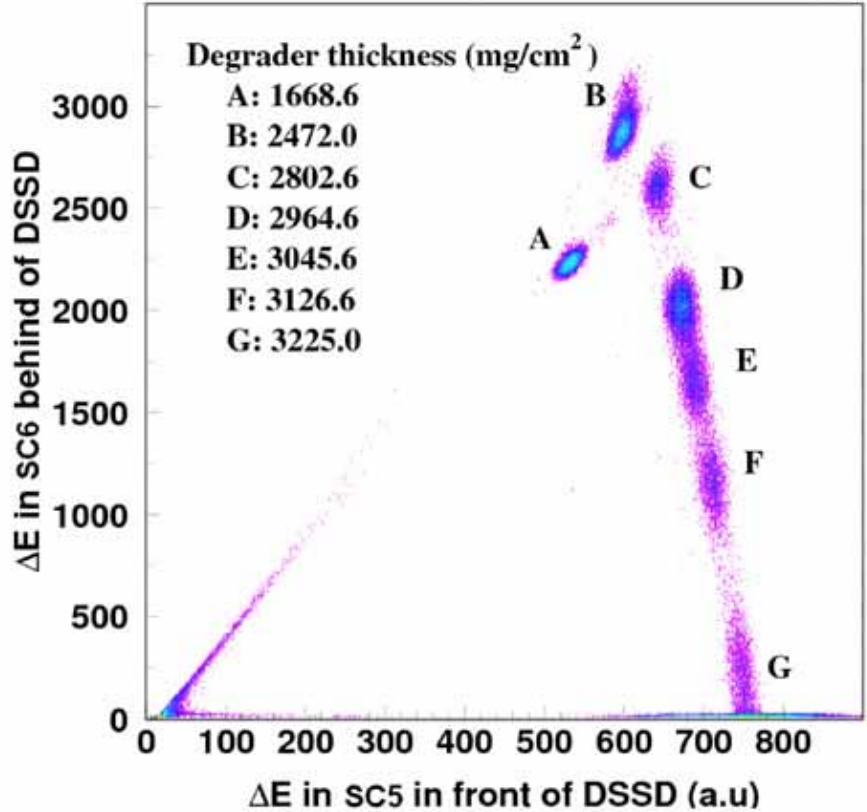

Figure 4: Calibration of the thickness of the aluminium degrader used during the experiment. The energy loss of the $^{208}Pb$ beam at the scintillator detectors placed both in front of and behind the catcher is represented. The different letters indicate the beam distributions corresponding to the different degrader thicknesses.

Fragment implantation events were established by software, first requiring a signal above the threshold in the scintillator SC5 in front of the active stopper, no signal in the veto scintillation detector SC6, and a high-energy signal in a given pixel $(x,y)$, where $x$ is a horizontal strip and $y$ a vertical strip of the DSSD (see Figure 3). The particle identification data $(Z,A)$, as well as the implantation time, were stored. A subsequent β-like event in the same pixel was the one that produced a signal above the threshold in a single strip in both front and back of the DSSD.

The essential difference from previous experiments is that the primary beam has a time structure which interferes with the time-dependent decay rate and which also produces a background with a structure in time. Until now, the experiments have always been performed avoiding this complication by (a) choosing a time structure which can be approximated either by a delta function in time or (b) by providing a continuous production. The first case can be attained by choosing beam pulses which are short compared to the half-life to be measured and a repetition rate of beam pulses which is low compared to the inverse of the expected half-life. Also the suppression of the beam after the production of a nucleus falls into this category. The second case can only be attained if the accelerator is able to provide a continuous beam. A synchrotron, however, has to be filled, and the projectiles have to be accelerated repeatedly. Thus, periodic interruptions of the beam cannot be avoided. Still, one can approximate the conditions of a continuous beam if the length of the beam pulses is long compared to the half-life of the nucleus to be determined.

In our case, there are severe disadvantages, if one wants to achieve case (a). In particular a low repetition rate leads to a considerable loss of counting statistics, if the accelerator has to be driven with a repetition rate, which is lower than the accelerator is able to deliver. At SIS18, the highest possible repetition rate is about one per several seconds, depending on the nature and the end energy of the beam. Case (b) finds its limit in the maximum extraction length due to technical reasons, which is about 10 s. There are two other disadvantages connected with a long extraction time: firstly, a lowering of the beam dose achievable in a given time and, secondly, severe background conditions due to the continuous presence of the primary beam. In any case, the adaptation of the conditions requires a good knowledge of the half-life in advance. If the predictions are largely uncertain, an optimization of the experimental parameters is impossible.

Therefore, facing this experimental challenge, we have chosen an alternative solution: We abandon the requirement that the data analysis can be performed on the basis of an analytical description and thus gain considerably more freedom in optimizing the experimental conditions with respect to the final result and its uncertainty with the need to respect the boundary conditions given by the accelerator operation. This solution requires the development of an analysis procedure which copes with the rather complex time structures present in the accumulated decay-time distributions.

## 5. Numerical analysis method

Background is an essential parameter for the analysis of the time-interval sequences between implantations and β-decays. In order to be able to disentangle the background from the true β-decays in our approach, it is necessary to establish the time distribution of the uncorrelated events. This distribution is evaluated by histogramming the time difference between a given implantation to a previous β-like event, that is, making the fragment-β correlations in a time-reversed sequence. The ratio between the forward- and backward-time correlations contains the information on the 'true' fragment-β correlations.

The rather complex background found in the present experiment cannot be easily modelled with an *analytical* function. Our proposition is to determine the half-lives by using a *numerical* function to calculate the number of events per time interval. This function is obtained from Monte-Carlo simulations of time correlations between implantations and β-like events. The half-lives of the different nuclides under study were obtained from two-dimensional fits of the measured and simulated ratios of time correlations in forward- and backward-time direction. This was achieved by applying the least-squares method in which the lifetime τ and the β-detection efficiency ε were the two fitting parameters.

The experimental time-correlation spectra correspond to a limited number of events and, hence, the number of counts in the different time bins is subject to statistical fluctuations according to the Poisson statistics. Therefore, the measured time-correlation spectrum will deviate from the simulated one, even if the parameters of the simulation were `correct'. In order to extract the parameters of interest, e.g. the half-life of the nuclide considered, we apply the chi-squared ($\chi^2$) method by determining the parameters of the simulation, which are most compatible with the measured spectra. In the $\chi^2$ test we assume Gaussian statistics and quote the standard deviation σ in a result from the Gaussian probability density function, in which approximately 68.3 % of the events of the Gaussian distribution fall within ±σ of the mean.

According to the least-squares method, the optimum values of the parameters $a_j$ of the fitting functions are obtained by minimising $\chi^2$ simultaneously with respect to all parameters. As the condition for minimising $\chi^2$ is that the first partial derivative with respect to each parameter cancels out (i.e., $\partial \chi^2 / \partial a_j = 0$), we can expect that near a local minimum with respect to any parameter $a_j$, $\chi^2$ will be a quadratic function of that parameter.

In addition, we can estimate the errors in the fitting parameters $a_j$ by varying each parameter around the minimum to increase $\chi^2$ by one from the minimum value. The contour plot of $\chi^2$ as a function of two parameters $a_1$ and $a_2$ is generally approximately elliptical near the minimum. The degree of correlation between the parameters is indicated by the tilt of the ellipse. In order to determine a confidence interval, that is, a region of the $a_1$-$a_2$ space in which we estimate there is for example, a 68 % probability of finding the true values of the two parameters we should consider the full range of the $\Delta\chi^2 = 1$ contour. This is equivalent to allowing $a_2$ to assume its best values for each chosen value of $a_1$. Similarly, if we wish to find the limits of two standard deviations, we should find the limits on the $\Delta\chi^2 = 4$ contour.

### 5.1 Monte-Carlo simulation
A Monte Carlo code has been developed to simulate the time sequences in the experiment. In this code, the time sequences of fragment implantation and β detection are simulated according to the experimental conditions (spill sequence, fragment implantation rate during the spill and background rate during spill and pause), leaving two free parameters: the β lifetime τ and the efficiency ε for the detection of the β decays of the nuclide of interest. The code produces time-correlation spectra in forward- and backward-time direction.

For simplicity, only one 'cell' is considered. This cell corresponds to the strip or pixel which was triggered by one implantation and by one β-like event. All frequencies given as input parameters refer to this cell. A realistic simulation of the experimental conditions with different counting rates in the different cells can be performed by an accumulation of a series of calculations in which the appropriate parameters for each cell are specified.

When $N_F$ nuclides are implanted, the expected number of β-like events in the time-correlation spectrum, recorded in a time interval $[t, t + \Delta t]$ is $\Delta N(t) = N_F \cdot \rho(\tau, \varepsilon, t) \cdot \Delta t$ where $\rho(\tau, \varepsilon, t)$ is the probability density for detecting a β-like event in a time $t$ after the implantation of the considered nuclide.

This β-like event can be a 'true' β coming from the decay of the nuclide under study or a background event. The analysis can be performed on all consecutive β-like events detected in a given time interval or on the first β-like event detected following the implantation.

**5.2 Fitting procedure**
The rates of β-like events found during the beam spill and in the pause between spills differ highly from one implantation setting to another, being in general much higher during the spill than in the pause. This is due to high production rate of some species and the presence of δ electrons produced by the fast highly charged ions. As a consequence, when correlating with the first β-like event, if the frequency of β-like events during spill is too high, we will perform the time correlations mostly during the spill, losing the information of the βs that decay during the beam pause. In order to overcome this problem, the analysis of time correlations can be performed avoiding the beam pulses, that is, making the time correlations only in the pause between spills.

As an example in Figure 5 and Figure 6 we show the corresponding time-correlation spectra in forward- and backward-time directions, making the time correlations of the implantation and the first consecutive β-like event in both spill and pause and only during the pause, respectively, (upper panels) and an example of the same distributions simulated with the Monte Carlo code (bottom).

The data correspond to $^{195}$Re, a heavy neutron-rich nuclide synthesised for the first time in this experiment. We select this nuclide because the experimental conditions allow us to study the fragment – β-time correlations in both options, with the full time distribution and only during the pause.

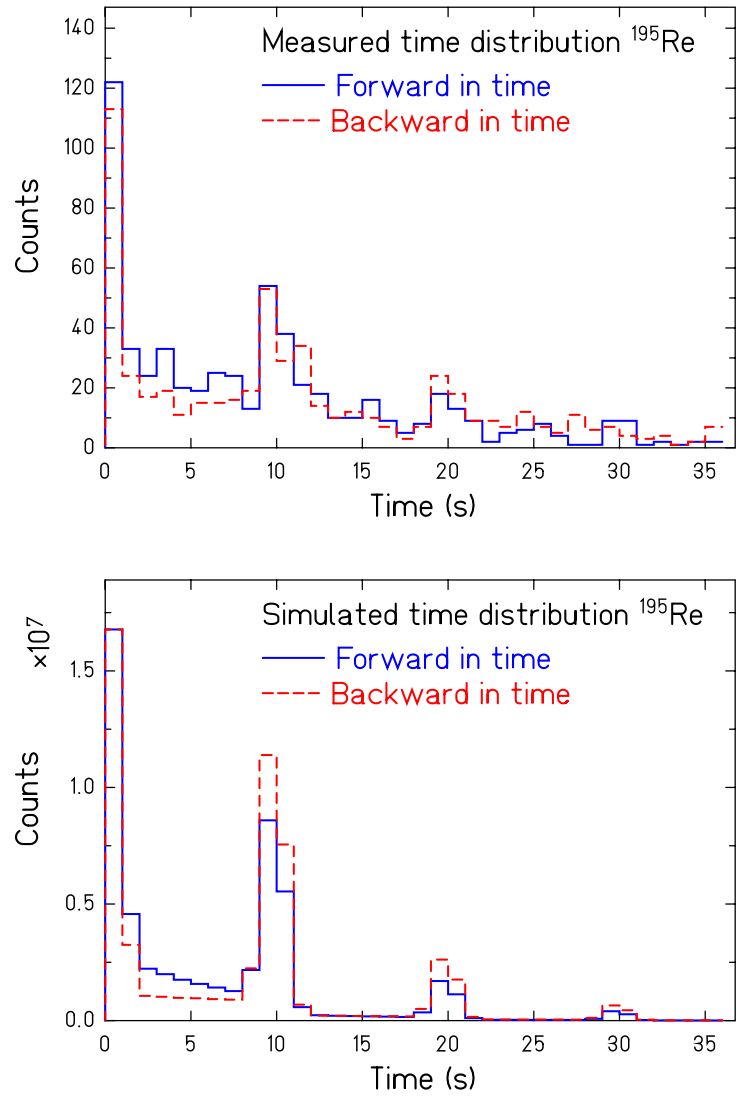

Figure 5: Measured implantation-β forward- and backward-time distributions for $^{195}$Re for the full time structure (spill+pause) (top) and an example of the same distributions simulated with a Monte Carlo code assuming $\tau = 8$ s and 40 % β-detection efficiency (bottom).

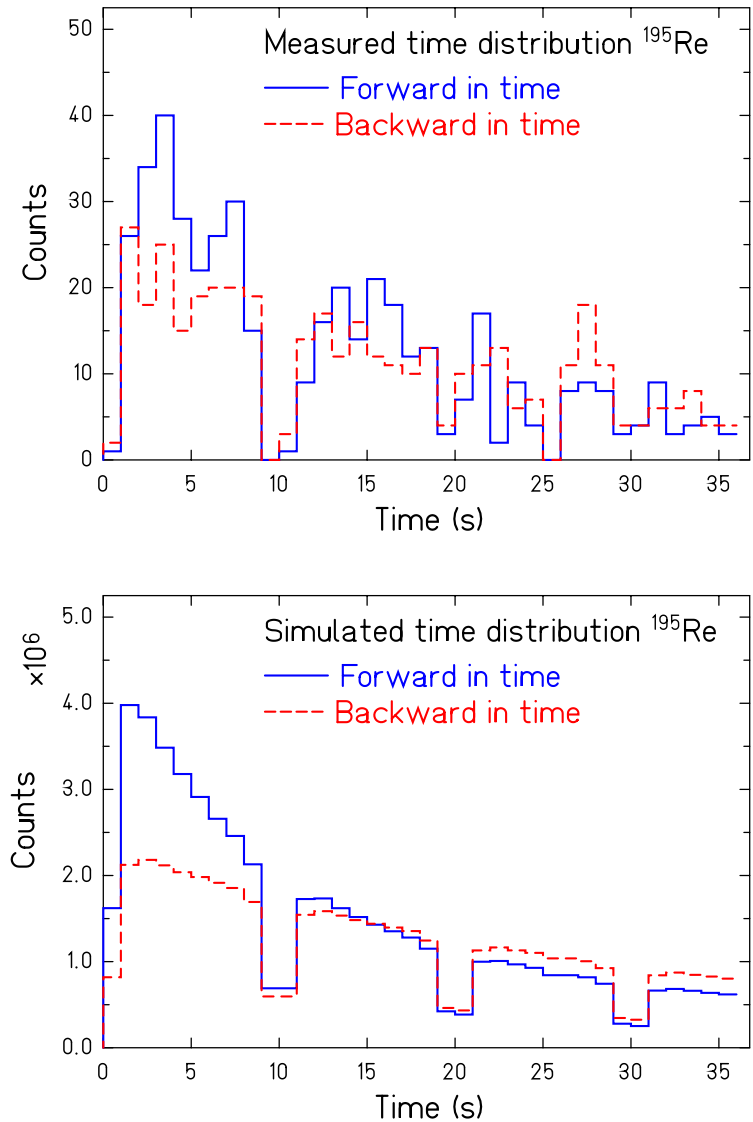

Figure 6: Measured implantation-β forward- and backward-time distributions for $^{195}$Re for the time correlations performed only during the pause between beam pulses (top) and an example of the same distributions simulated with a Monte-Carlo code assuming τ = 9 s and 30 % β-detection efficiency (bottom).

Figure 7 shows the measured forward/backward ratios with the experimental error bars, and a comparison with three different Monte-Carlo simulated ratios, for a fixed efficiency and different lifetimes in both, only during the beam pauses (top) and using the full time structure (spill+pause) (bottom). Note that the simulations have been performed with high statistics, imposing that the statistical uncertainties of the simulations are negligible. From these graphs we learn that it is necessary to make time correlations using several spills in order to distinguish between different lifetimes.

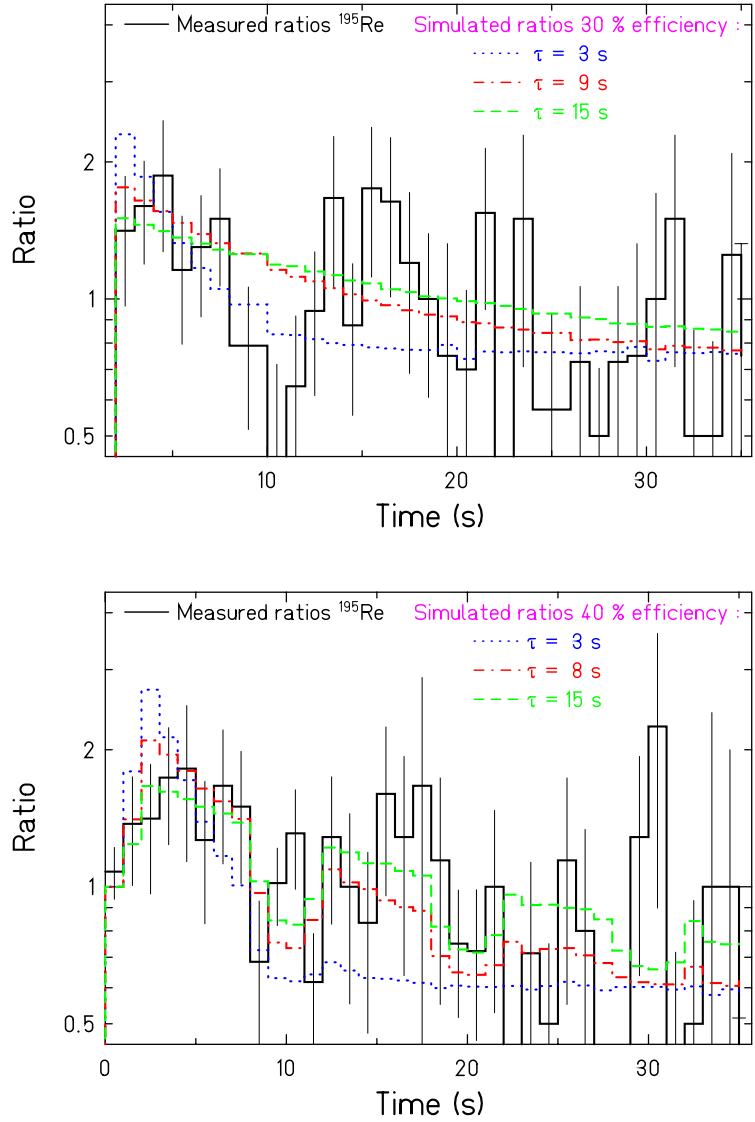

Figure 7: Top: Ratios of the time-difference spectra between the implantation of $^{195}$Re and the first β-like particle detected in the same strip during the pause between beam pulses in forward- and backward-time and the corresponding Monte-Carlo-simulated ratios for different lifetimes and an efficiency of 30 %. Bottom: ratios of the time distribution using the full time structure compared with Monte Carlo simulated ratios for different lifetimes and ε = 40 %.

In order to determine the half-life of the selected nuclide, we perform sets of simulations with a given efficiency and lifetime, and we calculate the $\chi^2$ from the measured and simulated ratio of the spectra of time correlations in forward- and backward-directions for each set of simulations. This allows us to perform 2-dimensional $\chi^2$ contour plots like the ones shown in Figure 8. From these contour plots we found that there is a minimum in both parameters, efficiency and lifetime, from which we can determine the half-life of the nuclide by fitting the region around the minimum of the $\chi^2$ distribution to a parabola. Both types of analysis (full time and only pause) give consistent results within the error bars, that is, $T_{1/2} = (6 \pm 2)$ s when using the full time structure, and $T_{1/2} = (6 \pm 1)$ s when correlating fragment and β-decays only between beam pulses. The analysis only in the pause gives smaller error bars in this case, which is consistent with the fact that by avoiding the beam pulses the time correlations are less dominated by the background. This gives us confidence in the results of the half-lives determined in the cases when it is only possible to study the time-correlation distributions in the pause between the beam pulses (due to the high β background present during the spill).

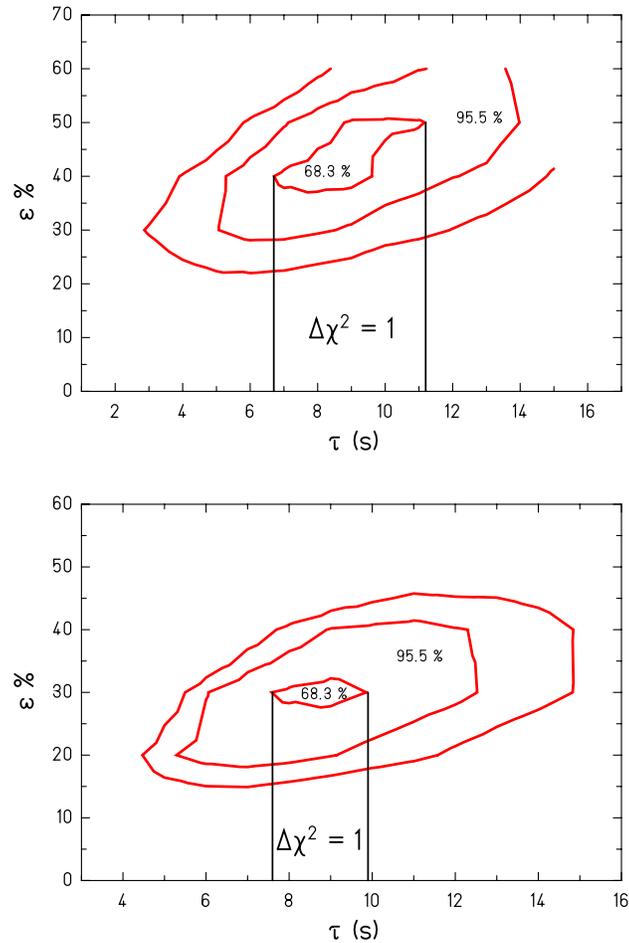

Figure 8: Top: $\chi^2$ two-dimensional contour plot (lifetime-efficiency) obtained with our fitting procedure when considering the full time structure, spill and pause, and (bottom) the corresponding one for only the pause between spills.

In order to test the method, a measurement of the half-life of $^{198}$Ir was used, because its half life has been determined previously to 8 s [10,11]. However, it is important to note that the conditions of our experiment for this nucleus were less favourable than for $^{195}$Re. As discussed above, background-rate values differ from one FRS setting to another, due to variation in the production rate of the implanted species in the different settings and the overall build-up in activity over the course of the experiment. In particular, for the $^{198}$Ir FRS setting, different isotopes of Pt, Ir and Os were implanted, which were produced with a high rate, allowing short time windows ≤ 30 s, and also there was an important contribution of δ electrons, which prevented us from performing time-correlation distributions using the full time range (spill+pause) but only in the pause between beam pulses. However, this is not an important constraint since the expected half-life of the nuclide is 8 s, longer than the length of the beam pulse, and, as shown for $^{195}$Re, half-lives determined using the time-correlation distributions only in the pause between the beam pulses are less dominated by the background.

Figure 9 shows the distribution of the minimum $\chi^2$ in relation to the lifetime parameter, corresponding to $^{198}$Ir. The minimum $\chi^2$ of the fit for the measured and simulated ratio of the forward- and backward-time correlations corresponds to $T_{1/2}$ = 8 s ± 2 s which is in perfect agreement with the previous measurement.

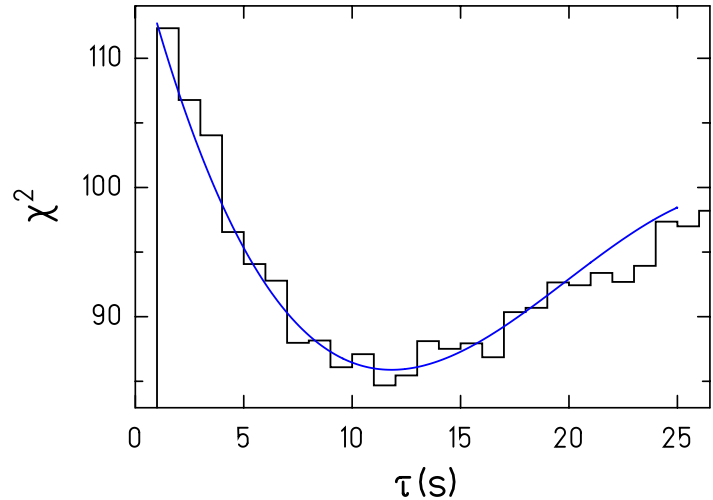

Figure 9: $\chi^2$ projection onto the lifetime coordinate for a 40 % detection efficiency for $^{198}$Ir. The smooth line shows the fit with a third-order polynomial.

### 5.3 Applicability of the method
The analysis method described before assumes Gaussian statistics in the $\chi^2$ calculation. As a consequence, the number of counts in the histogram bins has to be approximately > 10 [5]. In this case, the Poisson distribution can be well approximated by a Gaussian function.

Following the discussion on the statistical analysis presented in Ref. [12] and Ref. [13], let $N_F$ be the total number of implanted fragments from which we can measure the half-life $T_{1/2} = \tau \cdot \ln 2$, $\nu_\beta$ the β-background frequency measured per detector, and ε the β detection efficiency. The number of 'true' β-decays detected during a time equal to $T_{1/2}$ after the implantation of the fragment, is then $N_F \, \varepsilon/2$, and the β-like background detected in the same time is $N_F \, \nu_\beta \, T_{1/2}$. The fluctuation of the random events at the same time is the square root of the last formula.

From the experience gained during the analysis we found that we can expect to obtain reliable results with our analysis method, if the number of true correlations is at least four times the number of random correlations, leading to the condition

$$\frac{N_F \varepsilon}{2} > 4\sqrt{N_F \nu_b T_{1/2}}$$

and then we obtain an upper limit of the value of the half-life that we can measure

$$T_{1/2} < \frac{N_F \varepsilon^2}{64 \nu_\beta}$$

Two parameters that constrain the time intervals in the histograms are the maximum correlation time $T_c$ and the minimum time between implanted ions. The maximum correlation time is defined as the longest time window in which a β-decay can be correlated with an implanted ion. The time $T_c$ needs to be long enough to encompass between 2 to 5 times the expected value for the half-life [14] of the nuclide under study in order to extract a reasonable decay curve. Longer $T_c$ times would increase the chance of correlating the implanted ion of interest with a background event. The minimum implantation time is defined as the minimum average time span between two successive implantation events. The time window defined by the minimum implantation time depends on the implantation probability in the same detector. The implantation probability follows a Poisson distribution and can be written as follows:

$$P(t) = \frac{\nu_t \cdot t^n}{n!} \cdot e^{-\nu \cdot t}$$

where ν is the implantation rate, $t$ is the time window of the measurement and $n$ is the number of nuclides implanted in the same detector. In order to calculate the probability of multiple implantation in the same detector we need to determine the implantation probability of more than 1 nuclide, that is, $P_m = 1 - P_0 - P_1$, where $P_0$ is the probability of implantation of no one nuclide and $P_1$ is the implantation probability of only one nuclide. As a consequence, depending on the implantation rate different time windows will be available for the study of the different nuclides.

## 6. Choice of the correlation option

The spectrum of time intervals between the implantation of a given nuclide and the following β decay can be analysed in different ways, e.g. either correlating all consecutive β-like events detected within a given time window or only the first β-like event detected after implantation. A comparison of these two options was first discussed in Ref. [13]. In that study, the experimental half-life values obtained using both methods were found to be statistically compatible, and the final adopted values for half-lives were taken by averaging the results of the two methods.

The analysis of the present data discussed in the previous chapters relies on the correlation with the first beta-like event after implantation, but we have also studied the correlation with all consecutive beta-like events. Using the second option, we found that the influence of statistical uncertainties on the analysis was appreciably stronger, preventing even to extract the lifetime of the implanted nuclides. We understand this finding in the following way: When the correlation to all consecutive beta-like events is established, the spectrum contains the fraction of true betas given by the detection efficiency plus *all* background events in the time window up to the next implantation. If the background is high, the number of counts in this spectrum is strongly dominated by the background events. That means that most of the counts in the spectrum are not significant for the lifetime of the nucleus. When only the first consecutive beta-like event is recorded, the numbers of both the true decays and the background events are reduced. The true decays are reduced by the probability to register a background event before the true beta decay. However, the background contribution to the correlation spectrum is reduced from a continuous sequence to at most one event. It is difficult to draw a general conclusion from these considerations, in particular in the case of the complex time structure of our experiment, but it appears plausible that the influence of statistical fluctuations of background events on the analysis is weaker if only the first beta-like event is recorded.

In previous applications of the delayed-coincidence method, e.g. [13,15,16], the importance of the correlation option for the influence of the background rate on the analysis did not show up so clearly, because the uncertainties were mostly determined by the limited statistics of the observed events and not by the contribution of background.

It is obvious that more than the first β-like event following the implantation has to be analyzed, if one is interested in correlating the β decay of the implanted nucleus with the β decays of daughter nuclei, as it was done in refs. [15,16]. Daughter decays are strongly shielded if only the first β-like event is analyzed.

# 7. Conclusion

The present work represents a first step towards the study of heavy neutron-rich nuclei, which are important for understanding the astrophysical r-process and to benchmark nuclear models far from stability. The progress relies on the implantation of projectile-like reaction products obtained by cold fragmentation in an active catcher. Due to the specific experimental conditions of this technique, conventional analysis tools, based on analytical time-distribution functions could not be applied. Therefore, a new analysis procedure has been developed to extract the β-decay half-lives of the exotic nuclei, and its success has been demonstrated. The new method copes with the complex time structure present in the time-interval distributions of the fragment-β correlations due to the periodic operation cycles of the heavy-ion synchrotron, providing the primary beam. We have shown that this time structure of the secondary beams can be used directly without losses due to additional beam suppression. The new method is generally suited to analysing experimental time distributions which are modulated with periodic source variations.

In addition, we have found that recording only the first decay-like event in the present experiments reduces the influence of background compared to recording all decay-like events in a given time interval. This is an important finding for the analysis of experiments with high background rates.

**Acknowledgements**

This work was supported by the Spanish MEC (FPA2005-00732), Xunta de Galicia (Consellería de Educación) and by the European Commission (EURONS, contract no. 506065). I.N.B acknowledges the support by German DFG under the contract 436 RUS 113 907/0-1. T.K.-N. acknowledges the support from Xunta de Galicia, Dirección Xeral de Investigación, Desenvolvemento e Innovación (Bolsa Predoutoral, Consellería de Innovación e Industria) and Universidade de Santiago de Compostela, through the PhD studies.